\documentstyle[twoside]{article}
\catcode`\@=11
\long\def\@makefntext#1{
\protect\noindent \hbox to 3.2pt {\hskip-.9pt
$^{{\eightrm\@thefnmark}}$\hfil}#1\hfill}               %CAN BE USED

\def\thefootnote{\fnsymbol{footnote}}
\def\@makefnmark{\hbox to 0pt{$^{\@thefnmark}$\hss}}    %ORIGINAL

\def\ps@myheadings{\let\@mkboth\@gobbletwo
\def\@oddhead{\hbox{}
\rightmark\hfil\eightrm\thepage}
\def\@oddfoot{}\def\@evenhead{\eightrm\thepage\hfil
\leftmark\hbox{}}\def\@evenfoot{}
\def\sectionmark##1{}\def\subsectionmark##1{}}

\oddsidemargin=\evensidemargin
\renewcommand{\thefootnote}{\fnsymbol{footnote}}

\newcounter{sectionc}\newcounter{subsectionc}\newcounter{subsubsectionc}
\renewcommand{\section}[1] {\vspace{12pt}\addtocounter{sectionc}{1}
\setcounter{subsectionc}{0}\setcounter{subsubsectionc}{0}\noindent
        {\tenbf\thesectionc. #1}\par\vspace{5pt}}
\renewcommand{\subsection}[1] {\vspace{12pt}\addtocounter{subsectionc}{1}
        \setcounter{subsubsectionc}{0}\noindent
        {\bf\thesectionc.\thesubsectionc. {\kern1pt \bfit #1}}\par\vspace{5pt}}
\renewcommand{\subsubsection}[1] {\vspace{12pt}\addtocounter{subsubsectionc}{1}
        \noindent{\tenrm\thesectionc.\thesubsectionc.\thesubsubsectionc.
        {\kern1pt \tenit #1}}\par\vspace{5pt}}
\newcommand{\nonumsection}[1] {\vspace{12pt}\noindent{\tenbf #1}
        \par\vspace{5pt}}

\newcounter{appendixc}
\newcounter{subappendixc}[appendixc]
\newcounter{subsubappendixc}[subappendixc]
\renewcommand{\thesubappendixc}{\Alph{appendixc}.\arabic{subappendixc}}
\renewcommand{\thesubsubappendixc}
        {\Alph{appendixc}.\arabic{subappendixc}.\arabic{subsubappendixc}}

\renewcommand{\appendix}[1] {\vspace{12pt}
        \refstepcounter{appendixc}
        \setcounter{figure}{0}
        \setcounter{table}{0}
        \setcounter{lemma}{0}
        \setcounter{theorem}{0}
        \setcounter{corollary}{0}
        \setcounter{definition}{0}
        \setcounter{equation}{0}
        \renewcommand{\thefigure}{\Alph{appendixc}.\arabic{figure}}
        \renewcommand{\thetable}{\Alph{appendixc}.\arabic{table}}
        \renewcommand{\theappendixc}{\Alph{appendixc}}
        \renewcommand{\thelemma}{\Alph{appendixc}.\arabic{lemma}}
        \renewcommand{\thetheorem}{\Alph{appendixc}.\arabic{theorem}}
        \renewcommand{\thedefinition}{\Alph{appendixc}.\arabic{definition}}
        \renewcommand{\thecorollary}{\Alph{appendixc}.\arabic{corollary}}
        \renewcommand{\theequation}{\Alph{appendixc}.\arabic{equation}}
        \noindent{\tenbf Appendix \theappendixc #1}\par\vspace{5pt}}
\newcommand{\subappendix}[1] {\vspace{12pt}
        \refstepcounter{subappendixc}
        \noindent{\bf Appendix \thesubappendixc. {\kern1pt \bfit #1}}
        \par\vspace{5pt}}
\newcommand{\subsubappendix}[1] {\vspace{12pt}
        \refstepcounter{subsubappendixc}
        \noindent{\rm Appendix \thesubsubappendixc. {\kern1pt \tenit #1}}
        \par\vspace{5pt}}

\newcommand{\textlineskip}{\baselineskip=13pt}
\newcommand{\smalllineskip}{\baselineskip=10pt}

\def\eightcirc{
\begin{picture}(0,0)
\put(4.4,1.8){\circle{6.5}}
\end{picture}}
\def\eightcopyright{\eightcirc\kern2.7pt\hbox{\eightrm c}}

\newcommand{\copyrightheading}[1]
        {\vspace*{-2.5cm}\smalllineskip{\flushleft
        {\footnotesize Modern Physics Letters A, LA-UR-98-2905 and EFUAZ FT-98-62}\\
        {\footnotesize $\eightcopyright$\, World Scientific Publishing
         Company}\\
         }}

\def\abstracts#1#2#3{{
        \centering{\begin{minipage}{4.5in}\baselineskip=10pt\footnotesize
        \parindent=0pt #1\par
        \parindent=15pt #2\par
        \parindent=15pt #3
        \end{minipage}}\par}}

\def\keywords#1{{
        \centering{\begin{minipage}{4.5in}\baselineskip=10pt\footnotesize
        {\footnotesize\it Keywords}\/: #1
         \end{minipage}}\par}}

\newcommand{\bibit}{\nineit}
\newcommand{\bibbf}{\ninebf}
\renewenvironment{thebibliography}[1]
        {\frenchspacing
         \ninerm\baselineskip=11pt
         \begin{list}{\arabic{enumi}.}
        {\usecounter{enumi}\setlength{\parsep}{0pt}
         \setlength{\leftmargin 12.7pt}{\rightmargin 0pt} %FOR 1--9 ITEMS
         \setlength{\itemsep}{0pt} \settowidth
        {\labelwidth}{#1.}\sloppy}}{\end{list}}

\newcounter{itemlistc}
\newcounter{romanlistc}
\newcounter{alphlistc}
\newcounter{arabiclistc}

\newcommand{\fcaption}[1]{
        \refstepcounter{figure}
        \setbox\@tempboxa = \hbox{\footnotesize Fig.~\thefigure. #1}
        \ifdim \wd\@tempboxa > 5in
           {\begin{center}
        \parbox{5in}{\footnotesize\smalllineskip Fig.~\thefigure. #1}
            \end{center}}
        \else
             {\begin{center}
             {\footnotesize Fig.~\thefigure. #1}
              \end{center}}
        \fi}

\newcommand{\tcaption}[1]{
        \refstepcounter{table}
        \setbox\@tempboxa = \hbox{\footnotesize Table~\thetable. #1}
        \ifdim \wd\@tempboxa > 5in
           {\begin{center}
        \parbox{5in}{\footnotesize\smalllineskip Table~\thetable. #1}
            \end{center}}
        \else
             {\begin{center}
             {\footnotesize Table~\thetable. #1}
              \end{center}}
        \fi}

\def\@citex[#1]#2{\if@filesw\immediate\write\@auxout
        {\string\citation{#2}}\fi
\def\@citea{}\@cite{\@for\@citeb:=#2\do
        {\@citea\def\@citea{,}\@ifundefined
        {b@\@citeb}{{\bf ?}\@warning
        {Citation `\@citeb' on page \thepage \space undefined}}
        {\csname b@\@citeb\endcsname}}}{#1}}

\newif\if@cghi
\def\cite{\@cghitrue\@ifnextchar [{\@tempswatrue
        \@citex}{\@tempswafalse\@citex[]}}
\def\citelow{\@cghifalse\@ifnextchar [{\@tempswatrue
        \@citex}{\@tempswafalse\@citex[]}}
\def\@cite#1#2{{$\null^{#1}$\if@tempswa\typeout
        {IJCGA warning: optional citation argument
        ignored: `#2'} \fi}}

\def\@refcitex[#1]#2{\if@filesw\immediate\write\@auxout
        {\string\citation{#2}}\fi
\def\@citea{}\@refcite{\@for\@citeb:=#2\do
        {\@citea\def\@citea{, }\@ifundefined
        {b@\@citeb}{{\bf ?}\@warning
        {Citation `\@citeb' on page \thepage \space undefined}}
        \hbox{\csname b@\@citeb\endcsname}}}{#1}}

\def\@refcite#1#2{{#1\if@tempswa\typeout
        {IJCGA warning: optional citation argument
        ignored: `#2'} \fi}}

\def\refcite{\@ifnextchar[{\@tempswatrue
        \@refcitex}{\@tempswafalse\@refcitex[]}}

\def\pmb#1{\setbox0=\hbox{#1}
        \kern-.025em\copy0\kern-\wd0
        \kern.05em\copy0\kern-\wd0
        \kern-.025em\raise.0433em\box0}

\def\fnt#1#2{\footnotetext{\kern-.3em
        {$^{\mbox{\scriptsize #1}}$}{#2}}}

\def\fpage#1{\begingroup
\voffset=.3in
\thispagestyle{empty}\begin{table}[b]\centerline{\footnotesize #1}
        \end{table}\endgroup}

\def\runninghead#1#2{\pagestyle{myheadings}
\markboth{{\protect\footnotesize\it{\quad #1}}\hfill}
{\hfill{\protect\footnotesize\it{#2\quad}}}}
\font\tenrm=cmr10
\font\tenit=cmti10
\font\tenbf=cmbx10
\font\bfit=cmbxti10 at 10pt
\font\ninerm=cmr9
\font\nineit=cmti9
\font\ninebf=cmbx9
\font\eightrm=cmr8

\def\qed{\hbox{${\vcenter{\vbox{                        %HOLLOW SQUARE
   \hrule height 0.4pt\hbox{\vrule width 0.4pt height 6pt
   \kern5pt\vrule width 0.4pt}\hrule height 0.4pt}}}$}}

\renewcommand{\thefootnote}{\fnsymbol{footnote}}        %USE SYMBOLIC FOOTNOTE

\begin{document}

%------------------------------------------------------------------------------
\def\be{\begin{equation}}
\def\ee{\end{equation}}

%------------------------------------------------------------------------------

\runninghead{D. V. Ahluwalia}
{On reconciling  atmospheric, LSND,
$\ldots$}
\normalsize\textlineskip
\thispagestyle{empty}
\setcounter{page}{1}

\copyrightheading{}                     %{Vol. 0, No.0 (1992) 000--000}

\vspace*{0.88truein}

\fpage{1}

\centerline{\bf ON RECONCILING ATMOSPHERIC, LSND, AND SOLAR}
\vspace*{0.035truein}
\centerline{\bf   NEUTRINO-OSCILLATION DATA}

\vspace*{0.37truein}

\centerline{\footnotesize D. V. AHLUWALIA}
\vspace*{0.015truein}
\baselineskip=10pt

\centerline{\footnotesize\it
Physics Division (P-25), Mail Stop H-846}
\centerline{\footnotesize\it Los Alamos National
Laboratory, Los Alamos, NM 87545 USA}
\vspace*{0.015truein}
\baselineskip=10pt
\centerline{\footnotesize\it and}
\vspace*{0.015truein}
\baselineskip=10pt
\centerline{\footnotesize\it
Escula de Fisica, Univ. Aut. de Zacatecas, Apartado Postal C-580}
\vspace*{0.015truein}
\centerline{\footnotesize\it
Zacatecas, ZAC 98068, Mexico}
\vspace*{0.015truein}
\baselineskip=10pt
\centerline{\footnotesize\it E-mail: av@p25hp.lanl.gov}

%\vspace*{0.225truein}
%\publisher{(received date)}{(revised date)}

\vspace*{0.21truein}
\abstracts{
The $L/E$-flatness of the $e$-like events observed in the recent
atmospheric-neutrino data from Super-Kamiokande (SuperK) is
interpreted to reflect a new symmetry of the neutrino-oscillation
mixing matrix. From that we obtain an analytical set of constraints
yielding a class of mixing matrices of the property to
simultaneously fit both the SuperK and the LSND data. The resulting
mass squared difference relevant for the LSND experiment is found
as $0.3\,\,\mbox{eV}^2$. The discussed symmetry, e.g., carries the
nature that expectation values of masses for $\nu_\mu$ and
$\nu_\tau$ are identical. These considerations are purely data
dictated. 
A different framework is then applied to the solar
neutrino problem. 
It is argued that a single sterile neutrino is an unlikely candidate 
to accommodate
the data from the four solar neutrino experiments.
A scenario is discussed which violates CPT symmetry, and
favors the
$\nu_e$-$\overline\nu_e$ system to belong to the `self'-`anti-self'
charge conjugate construct in the $(1/2,0)\oplus (0,1/2)$
representation space, where the needed helicity flipping amplitudes
are preferred, rather than the usual Dirac, or Majorana,
constructs. In the presented framework the emerging SuperK data on
solar neutrino flux is reconciled with the  Homestake,  GALLEX, and SAGE 
experiments. This happens because  the former detects not only the
solar $\nu_e$ but also, at a lower cross section,  the oscillated
solar $\overline\nu_e$; while the latter are sensitive only to the
oscillation-diminished solar $\nu_e$ flux. A direct observation of
solar $\overline\nu_e$ by SNO will confirm our scenario. Finally,
we consider the possibility for flavor-dependent gravitational
couplings of neutrinos as emerging out of the noncommutativity of
the quantum operators associated with the measurements of energy
and flavor. }{}{}

\vspace*{10pt}

\vfill
\pagebreak

%\textlineskip                  %) USE THIS MEASUREMENT WHEN THERE IS
%\vspace*{12pt}                 %) NO SECTION HEADING

\vspace*{1pt}\textlineskip      %) USE THIS MEASUREMENT WHEN THERE IS

% Report No: EFUAZ FT-98-62, and LA-UR-98-2905

\section{Introduction}    %) A SECTION HEADING
\vspace*{-0.5pt}

\vbox{
\begin{quote}
{\bibbf ``Is it worth to search for the particle-antiparticle mass
differences in sectors other than $K^0\bar K^0$?'' L. B.
Okun.\cite{lo}}
\end{quote}
}

\noindent
With the preliminary results of Kamiokande and the LSND experiments
presented a few years ago,\cite{K,LSND} and now both groups
presenting more definitive evidence for neutrino
oscillations,\cite{dif,SK} the original Pontecorvo
suggestion\cite{bp} that the long-standing solar neutrino
anomaly\cite{Davis} may be pointing towards the repeat of the $K$
system in neutrino oscillations seems confirmed.

Evidence for neutrino oscillations provide empirical support for
flavor eigenstates of neutrinos  not to be mass eigenstates.
Instead, these flavor eigenstates are  suggested to be a linear
superposition of some underlying mass eigenstates
\be
\vert\nu_\ell\rangle = \sum_{j} \,U_{\ell j}\,
\vert m_j \rangle,\label{1}
\ee
where the flavor index  $\ell=e,\mu,\tau$ and mass index $j
=1,2,3$. The $\vert\nu_\ell\rangle$ and $\vert m_ j\rangle$
are flavor and mass eigenstates respectively, while $U_{\ell j}$
are elements of a $3\times 3$ unitary matrix to be determined from data.

Now, the question arises, why, after the $K^0\bar K^0$ system has
Nature chosen to create another physical system whose elements are
linear superposition of different mass eigenstates? May this be the
expression of something  fundamentally new? We here argue that this
fundamentally `new something' may be the violation of CPT symmetry,
and the related violation of the principle of equivalence.
Specifically, in this scenario the mass eigenstates underlying the
$\nu_e$ and $\overline\nu_e$ may carry slightly different masses.

These opening  remarks  are followed, in Section 2, by an
introduction to the SuperK data on the $e$-like events and the
questions raised by them. Section 3 is devoted to obtaining a set
of analytical constraints that the $L/E$-flatness of the e-like
events imposes. Section 4 contains an analytic and numeric study of
the constraint. This study yields a class of mixing matrices $U$.
This class of matrices is investigated in light of the existing
data to yield a subclass that fits all existing data except the
data on the solar neutrino deficit. The solar neutrino deficit
problem is then investigated separately in Sections 5,  6  and 7 where
considerations enumerated in the abstract are established. Section
8 closes the essay with a few concluding remarks. Unless otherwise
noted we follow the notation of Ref. [\refcite{notes}].
%\pagebreak

The reader whose interest is purely in the reconciliation of the
SuperK data with  the LSND experiment need only read Sections 2, 3,
and 4. These sections establish that, in the absence of CP
violation, the SuperK observed $L/E$-flatness of the e-like events
essentially determines the neutrino mixing matrix $U$.

Sections 5, 6 and 7 carry the flavor of  the `oral tradition.' There,
we suggest experimentally verifiable  implications of the emerging
data on the solar neutrino flux. These sections argue that the
apparent incompatibility between SuperK solar neutrino data and
Homestake chlorine experiment, GALLEX and SAGE
may be pointing towards a possible
new era in physics where CPT symmetry and the principle of
equivalence are only approximately valid.

Thus, this essay consists of two themes: (a)  The $L/E$-flatness of
the e-like events in the SuperK data severely restricts the class
of allowed neutrino mixing matrices; (b) One needs to be very
careful and open minded while looking at the apparent and emerging
incompatibilities between various solar neutrino experiments.

\textheight=7.8truein
\setcounter{footnote}{0}
\renewcommand{\thefootnote}{\alph{footnote}}

\section{Peculiarities  of the Super-Kamiokande data on the e-like events}

One of the noteworthy  results of the recent SuperK data on
atmospheric neutrinos is the  $L/E$ (i.e. zenith angle) dependence
of the following ratios
\begin{eqnarray}
{\cal R}_e&\equiv& \frac{\mbox{Experimentally observed e-like
events}}{\mbox{Theoretically expected e-like events (without $\nu$
oscillations)}},\\ {\cal R}_\mu&\equiv& \frac {\mbox{Experimentally
observed $\mu$-like events}} {\mbox{Theoretically expected
$\mu$-like
 events (without $\nu$ oscillations)}}
 .\label{result}
\end{eqnarray}
A remarkable feature of the SuperK data is that, while the second
ratio  ${\cal R}_\mu $ reveals a significant zenith angle
dependence, ${\cal R}_e$ is, within experimental errors,
$L/E$-independent and consistent with unity.\cite{SK} To be more
precise, slight deviations from the cited flatness and unity cannot
be ruled out. However, to gain first order insights into the
observed evidence for neutrino oscillations we shall assume the
flatness and  take it to be identical to unity.

Interpreting the zenith angle dependence of ${\cal R}_\mu $ and
${\cal R}_e$ as evidences for neutrino flavor oscillations, the
SuperK collaboration has tentatively concluded the maximal
$\nu_\mu\leftrightarrow
\nu_\tau
\,\,(\mbox{or}\,\,\nu_s)$
 mixing by taking the neutrino mixing
matrix to be of the form given below

\begin{equation}
U=\left(\begin{array}{ccc} 1 & 0 & 0\\ 0& c_\Theta & s_\Theta \\ 0
&
-s_\Theta & c_\Theta\end{array}\right),\label{usk}
\end{equation}
with $c_\Theta=\cos(\Theta)$ and $s_\Theta=\sin(\Theta)$, respectively.

The interpretation  presented by SuperK has the disadvantage  of
ignoring  the results from the LSND experiment, where the neutrino
source is apparently best understood, and which has reported a
direct evidence for neutrino oscillations in two
channels.\cite{LSND,dif}\footnote{
The  ``exclusion region''
presented recently by KARMEN experiment\cite{KARMEN}
appears to cover most of the LSND allowed parameter space. 
However, their exclusion curve lies well
outside the detector's sensitivity and emerges from zero 
events observed for an expected background of $2.8$ events.
I thank LSND collaboration for this observation.}

The independence of ${\cal R}_e$ on the $L/E$ parameter together
with its closeness with unity either implies, as SuperK inferred,
that the mixing matrix is given by Eq. (\ref{usk}), or, that there
is some underlying new symmetry hidden in the full $3\times 3$
space spanned by the three neutrino flavors.

By means of the scenario envisaged by SuperK, the $E$-dependence of
the solar neutrino deficit along with the atmospheric neutrino
anomaly could perhaps be explained (see Sec. 5 for the cautionary
tone of this remark). Still, the mixing matrix given by Eq.
(\ref{usk}) with a string of zeros seems too accidental. In the
zenith angle independence of ${\cal R}_e$ we suspect a hint on some
symmetry hidden in the mixing matrix, otherwise the flatness at
${\cal R}_e$
=1 would appear to be much too accidental.\footnote{
One may think of looking at the water level in a vessel. If one
sees no change in the water level one may assume no inflow, or
outflow, of water. Ordinarily, such flows would have to cancel each
other remarkably to yield a constant water level. However, there is
an exception. That case belongs to the situation when there is a
`water pump' connecting the two flows.  The zenith angle
independence of ${\cal R}_e (\simeq 1)$ poses a similar situation.
The neutrino mixing matrix (that acts as a counterpart of the water
pump), however, contains more than a single inflow and outflow
channel. Therefore, for the experimentally observed situation the
possibility that the neutrino mixing matrix carries certain
symmetries that cancel the inflow and outflow channels rises.}

>From that point of view we are going to study below the surprising
independence of the ${\cal R}_e(\simeq 1)$ ratio on the $L/E$
parameter. To simplify the mathematical structure of the analysis
and for the sake of transparency of the physical insights, we
henceforth use systematically the unit value for ${\cal R} _e$.
Accommodations can be considered later for any relatively small
variations (with $L/E$) in ${\cal R}_e$ that may emerge as more
data become available. The purpose at present is to roughly obtain
the structure of all
 mixing matrices compatible with the restriction
${\cal R}_e=1$ (considered as an equation) for all the relevant
$L/E$'s and use them to reveal compatibility between the various
neutrino oscillation experiments.

In other words, in a full three-flavor neutrino oscillation basis,
the independence of ${\cal R}_e$ on the $L/E$ parameter either
indicates that (a) the neutrino-mixing matrix be (\ref{usk}), thus
effectively reducing the mixing matrix  to a $2\times 2$ space, or
(b) for it to be ``natural'' the $L/E$-independence of ${\cal R}_e$
must arise from some very specific underlying class of mixing
matrices. If we discover this underlying class of matrices, we may
be able to unearth some yet-unknown symmetry hidden in neutrino
oscillations. At the same time this may allow us to consider an
alternative scenario which allows  the relevant
atmospheric-neutrino-anomaly $\Delta m^2
\simeq 0.6\times 10^{-2}\,\, \mbox{eV}^2$ and  the LSND-relevant $\Delta m^2
\simeq 0.3\,\, \mbox{eV}^2$ to explain all terrestrial experiments.
The $E$-dependence of the solar neutrino deficit would then require
either a sterile neutrino, or other possibilities as discussed in
Sections 5 and 6.

It is to be noted that for the atmospheric neutrino anomaly the
LSND-relevant mass squared difference  may play a significant role
for the data point corresponding to $L\sim 20\,\,\mbox{km}$, and
then that same mass squared difference  makes a constant
contribution for $L\gg 20\,\,\mbox{km}$. This happens because $L/E$
for LSND, and that for the  $L\sim 20\,\,\mbox{km}$ bin, carry
similar $L/E$ values. The precise contributions from the two mass
squared differences are determined, on a bin-by-bin basis, by the
various oscillation amplitudes determined by the mixing matrix.

\section{
Constraints on the neutrino-oscillation matrix as implied by  the
L/E-flatness of the e-like events}

In this section a general procedure for obtaining ${3\times 3}$
mixing matrices following from ${\cal R}_e=1$ condition is
developed.

At values of $t=0$, corresponding to the  ``top of the terrestrial
atmosphere,'' we will assume $N_e$ and $N_\mu $  to be the
respective numbers of $e$- and $\mu $-type neutrinos. Then within
the detector, i.e. at a distance $L= t$ away from the `top', the
number of $e$-type neutrinos will be
\begin{equation}
N^\prime_e = N_e P(\nu_e\rightarrow \nu_e) + N_\mu
P(\nu_\mu\rightarrow\nu_e).\label{5}
\end{equation}
Here, $P$ stands for the oscillation probability over the distance
$L$ at the relevant energy and in the indicated channel.\footnote{
It will be assumed that a relation that parallels Eq. (\ref{5}) is
valid for antineutrinos.}

Now, in assuming, for concreteness, $N_\mu/N_e = 2$ for the
relevant energy range, one easily realizes that the constancy of
${\cal R}_e=1$ over the range of $L/E$ relevant to SuperK implies
the following relation between the $\nu_e\to \nu_e$, and
$\nu_\mu\to
\nu_e $ oscillation probabilities:\footnote{If more precise data
were to establish the $L/E$ independent value of ${\cal R}_e$ to be
different from unity, then such a change can be  immediately
accommodated with minimal changes in the presented arguments.}
\begin{equation}
P(\nu_e\rightarrow \nu_e) + 2 P(\nu_\mu\rightarrow \nu_e) =1.
\label{one}
\end{equation}
A further constraint on the neutrino oscillation probabilities is
obtained from the unitarity condition on the $3\times 3$ neutrino
mixing matrix
\begin{equation}
U\,=\,
\left(
\begin{array}{ccccc}
c_\theta\,c_\beta &{\,\,}& s_\theta\,c_\beta &{\,\,}& s_\beta \\
-\,c_\theta\,s_\beta\,s_\psi\,e^{i\delta}-\,s_\theta\,c_\psi
&{\,\,}& c_\theta\, c_\psi\,-
e^{i\delta}\,s_\theta\,s_\beta\,s_\psi &{\,\,}& c_\beta\,s_\psi
\,e^{i\delta}\\
-\,c_\theta\,s_\beta\,c_\psi\,+\,s_\theta\,s_\psi\,e^{-i\delta}
&{\,\,}&
-\,s_\theta\,s_\beta\,c_\psi\,-\,c_\theta\,s_\psi\,e^{-i\delta}
&{\,\,}& c_\beta\,c_\psi
\end{array}\right),
\label{ucp}
\end{equation}
where $c_\beta\,=\,\cos(\beta)$, $s_\beta\,=\,\sin(\beta)$, etc.,.
as
\begin{equation}
P(\nu_e\rightarrow \nu_e) +  P(\nu_e\rightarrow \nu_\mu) +
P(\nu_e\rightarrow \nu_\tau)
 =1.\label{two}
\end{equation}
Allowing for a possible CP violation in the neutrino sector, Eqs.
(\ref{one}) and (\ref{two}) then yield:
\begin{equation}
2 P(\nu_\mu\rightarrow \nu_e) - P(\nu_e\rightarrow \nu_\mu) -
P(\nu_e\rightarrow \nu_\tau) =0.\label{theequation}
\end{equation}

Recall now the general expression for the
neutrino oscillation probabilities\cite{notes}
\begin{eqnarray}
P\left(\nu_\ell\rightarrow\nu_{\ell^\prime}\right) =
\delta_{\ell\,\ell'}
&-& \,4\,U_{\ell'\,1}\,U^\ast_{\ell\,1}\,U^\ast_{\ell'\,2}\,
U_{\ell\,2}\,\sin^2\left(\varphi^0_{2\,1}
\right)\nonumber\\
&-&\,4\,U_{\ell'\,1}\,U^\ast_{\ell\,1}\,U^\ast_{\ell'\,3}\,
U_{\ell\,3}\,\sin^2\left(
\varphi^0_{3\,1}\right)\nonumber \\
&-&\,4\,U_{\ell'\,2}\,U^\ast_{\ell\,2}\,U^\ast_{\ell'\,3}\,
U_{\ell\,3}\,\sin^2\left(
\varphi^0_{3\,2}\right)\,
,\label{eq:d}
\end{eqnarray}
 with the  kinematic phase being
 defined as\footnote{Note that the kinematic phases can be modified
for some dynamical reasons.\cite{msw,prd_ap98}}
\begin{equation}
\varphi^0_{\jmath\imath}
=
2\,\pi \,{L\over\lambda^{\rm osc}_{\jmath\imath }}.
\end{equation}
The definition of the oscillation length, $\lambda^{\rm osc}_{\jmath\imath }$,
is now standard and reads
\begin{equation}
\lambda^{{\rm {osc}}}_{\jmath
\imath}= {{2\,\pi}\over\alpha}\,{E\over{\Delta
m^2_{\jmath\imath}}}.
\end{equation}
The kinematic phase  may also be written as:
$\varphi^0_{\jmath\imath}
=1.27 \, \Delta m^2_{\jmath\imath } \times \left({L/ E}\right)$.
Here, $\alpha\,=\,\overline\alpha/2$; $\overline\alpha=2.54$ is
the usual factor that arises
from expressing $E$ in $\mbox{MeV}$, $L$ in meters, and $\Delta
m^2_{j\,k}$ in $\mbox{eV}^2\,$. $E$ refers to neutrino kinetic
 energy, $\sqrt{{\vec p}^{\,2}+m^2}$, and
$L$ is the distance between the creation region and the detection
region for the neutrino oscillation event. The six neutrino
oscillation parameters  are the two mass squared differences, the
three mixing angles, and a CP-violating phase angle $\delta$.

The $L/E$ independence of ${\cal R}_e =1 \,$, as translated into
Eq. (\ref{theequation}), now becomes a set of three coupled
equations

\begin{eqnarray}
&& 2\, U^\ast_{\mu\, 1} U_{\mu\,2} U_{e\,1} U^\ast_{e\,2}
   - U^\ast_{e\, 1} U_{e\,2} U_{\mu\,1} U^\ast_{\mu\,2}
   - U^\ast_{e\, 1} U_{e\,2} U_{\tau\,1} U^\ast_{\tau\,2} =0, \label{coupleda} \\
&& 2\, U^\ast_{\mu\, 1} U_{\mu\,3} U_{e\,1} U^\ast_{e\,3}
   - U^\ast_{e\, 1} U_{e\,3} U_{\mu\,1} U^\ast_{\mu\,3}
   - U^\ast_{e\, 1} U_{e\,3} U_{\tau\,1} U^\ast_{\tau\,3} =0, \label{coupledb}\\
&& 2\,   U^\ast_{\mu\, 2} U_{\mu\,3} U_{e\,2} U^\ast_{e\,3}
   - U^\ast_{e\, 2} U_{e\,3} U_{\mu\,2} U^\ast_{\mu\,3}
   - U^\ast_{e\, 2} U_{e\,3} U_{\tau\,2} U^\ast_{\tau\,3} =0.
\label{coupledc}
\end{eqnarray}

We thus have {\em three} transcendental equations in {\em four}
parameters of the mixing matrix $U$. These are the constraints on
the neutrino-oscillation matrices as implied by the $L/E$-flatness
of the $e$-like events.

Rigorously speaking, the considerations presented above should be
in terms of \cite{notes}
\be
{\cal{P}}(\nu_\ell\rightarrow\nu_{\ell^\prime}) =
\int_{E_{min}}^{E_{max}} dE \,f_\ell(E) \,P(\nu_\ell\rightarrow
\nu_{\ell^\prime}),
\ee
rather than in terms of $P(\nu_\ell\rightarrow\nu_{\ell\prime})$.
In the last equation  $f_\ell(E)$ is the neutrino flux as
normalized to unity, while $E_{min}$ and $E_{max}$ refer to the
minimum and maximum neutrino energy as determined by the combined
system of the beam and the detector. However, for the existing
precision of the data such a detailed analysis seems unwarranted.

\section{Analytical and numerical
solutions for neutrino-oscillation matrices}

Using the program-package for analytical calculations `Macsyma',
the left-hand sides of Eqs. (\ref{coupleda})-(\ref{coupledc}) can
be simplified towards expression containing only trigonometric
functions of a certain form. A dramatic simplification occurs on
setting $\psi=\theta=\pi/4$.  One is then led to the following set
of equations :
\begin{equation}
\left[ e^{-i\delta}\left(e^{i\delta}-1\right)
\left(e^{i\delta}+1\right)\right]c^2_\beta s_\beta =0,\quad
e^{i\delta} c^2_\beta s_\beta  = 0,\quad  e^{i\delta} c^2_\beta
s_\beta
= 0.\label{sol}
\end{equation}
Thus the first set of mixing angles $\{\theta=\pi/4, \beta=0,
\psi=\pi/4\}$ provides one possible  neutrino oscillation
matrix.\footnote{The set of angles $\{\theta=\pi/4, \beta=\pi/2,
\psi=\pi/4\}$, which also solves Eq. (\ref{sol}), does not result
in a full  $3\times 3$ mixing in the neutrino flavor space.} \,\,It
reads:

\begin{equation}
U=\frac{1}{\sqrt{2}}
\left(\begin{array}{ccccc}
1 &{\,\,}& 1 & {\,\,}& 0 \\
-{1}/{\sqrt{2}} &{\,\,}& {1}/{\sqrt{2}} &{\,\,}& e^{i\delta} \\
{e^{-i\delta}}/{\sqrt{2}} &{\,\,}& - {e^{-i\delta}}/{\sqrt{2}}
&{\,\,}& 1
\end{array}\right).\label{udetI}
\end{equation}
A second analytically obtained solution is found as:
\begin{equation}
U=\frac{1}{\sqrt{2}}
\left(\begin{array}{ccccc}
1&{\,\,}& 0 &{\,\,}& 1 \\
-e^{i\delta}/\sqrt{2} &{\,\,}& 1 &{\,\,}& e^{i\delta}/\sqrt{2} \\
-1/\sqrt{2}&{\,\,}& -e^{-i\delta}&{\,\,}& 1/\sqrt{2}
 \end{array}\right).\label{udetII}
\end{equation}

Finally, a set of physically irrelevant solutions has been found to
exist too. Those solutions  have the generic form in which each of
the three rows, at mutually non-intersecting positions, carries a $
+ 1$, or a $-1$, and two zeros in the remaining positions.

Apart from the analytical solutions presented above we also find
numerical similar solutions. As an example,
two typical numeric solutions (with the CP-violating phase angle $\delta$ being
set to zero) are given below as

\begin{equation}
{{{U}}\simeq\left(
\begin{array}{rrr  }
0.88 & 0.48 & 0.00 \\
-0.34 & 0.62 & 0.71 \\
0.34 & -0.62 & 0.71
\end{array}\right)},\quad
U\simeq\left(
\begin{array}{rrr  }
0.65 & 0.76 & 0.00 \\
-0.54 & 0.46 & 0.071 \\
0.54 & -0.46 & 0.071
\end{array}\right). \label{udetIII}
\end{equation}

These solutions  establish the existence of   mixing matrices in
the full $3\times 3$ dimensional neutrino flavor space so as to
satisfy the SuperK observed $L/E$ independence of ${\cal
R}_e(=1)$.\footnote{Note, how different $U$ of Eqs. (\ref{udetI}),
(\ref{udetII}), and (\ref{udetIII}) are from the SuperK's $U$ given
by Eq. (\ref{usk}).} However, none of these solutions satisfies the
other existing data and constraints. Indeed, assume a vanishing
CP-violating phase $\delta=0$ in agreement with the LSND's
decay-in-flight results.\cite{dif} Consider then, for concreteness,
the mixing matrix in Eq.~(\ref{udetII}). Since $U_{e 2}$
identically vanishes, the LSND relevant expression for the neutrino
oscillation probability simplifies to
\begin{eqnarray}
P(\overline{\nu}_\mu\rightarrow \overline{\nu}_e) &=&
-\,4\,U_{e\,1}\,U_{\mu\,1}\,U_{e\,3}\,
U_{\mu\,3}\,\sin^2\left(
\varphi^0_{3\,1}\right)\\
&=& 0.5\,
\sin^2\left(\frac {1.27\times
\Delta m^2_{3\,1} \times 30}{45}\right),
\label{eq:db}
\end{eqnarray}
where for  the LSND's decay-at-rest data we have set $\langle
L\rangle
\approx 30\,\,
\mbox{m}$, and $\langle E\rangle \approx 45\,\,
\mbox{MeV}$.
Setting LSND observed $P(\overline{\nu}_\mu\rightarrow
\overline{\nu}_e) \simeq 0.3\times 10^{-2}$, we obtain
$\Delta m^2_{3\,1}\simeq 0.1 \,\,\mbox{eV}^2$. One can further calculate
the probability for the disappearance of the electron antineutrino
(as in the reactor disappearance experiments) to obtain

\begin{eqnarray}
P(\overline{\nu}_e\rightarrow\overline{\nu}_e)&=&1-4\,
U_{e\,1}\,U_{e\,1}\,U_{e\,3}\, U_{e\,3}
\sin^2\left(\frac{1.27\times\Delta m^2_{3\,1} \times
L}{E}\right)\\ & =&1- 1.0 \,\sin^2\left(\frac{1.27\times\Delta
m^2_{3\,1}
\times L}{E}\right).
\end{eqnarray}
Taking $E\simeq 5 \,\,\mbox{MeV}$ as the typical reactor
$\overline{\nu}_e$ energy and setting $L=50 $m  and $L=100 $m
 for two representative settings of the detector in the
reactor $\overline{\nu}_e$ disappearance experiments, we obtain
$P(\nu_e\rightarrow\nu_e; L=50\,\,\mbox{m})
= 1- 0.91$, $P(\overline\nu_e\rightarrow\overline\nu_e; L=100\,\,\mbox{m}) = 1-
0.32$. These findings  are clearly in strong disagreement with the
known experimental results.\cite{reactor}

Finally, we also find numerical solutions of the type\footnote{The
essential difference with the above quoted solutions is in the
order-of-magnitude difference in the amplitude of oscillations
carried by each of the sine squared terms in the expression for
oscillation probability given by Eq. (\ref{eq:d}). Another related
matrix is:
\[
U \simeq
\left(\begin{array}{rrr}
0.99 & 0.10 & 0.00 \\
-\,0.07 & 0.70 & 0.70 \\
0.07 & -\,0.70 & 0.70 \end{array}\right).
\]
Also, now note the similarities and differences between these
solutions and that presented by SuperK -- cf. footnote {\bibit g}.}

\be\mbox{\boldmath$
U\simeq
\left(\begin{array}{rrr}
0.99 &  0.00 &  0.10 \\
 -\,0.07 & 0.71 &  0.70 \\
 -\,0.07 &  - \,0.71 &  0.70
\end{array}\right)$}. \label{ua}
\ee

These  mixing matrices have the property that the expectation values of the
masses associated with  the $\nu_\mu$ and $\nu_\tau$ neutrinos appear now
identical:
\begin{equation}
{\langle} m(\nu_\mu){\rangle} = {\langle} m(\nu_\tau) {\rangle}\,
.\label{exp}
\end{equation}
Here ${\langle} m(\nu_\ell){\rangle} \equiv U_{\ell 1}^2 m_1 +
U_{\ell 2}^2 m_2 +  U_{\ell 3}^2 m_3$.

Further, in using the mixing matrix in Eq. (\ref{ua}) in combination with
the LSND result, $P(\overline{\nu}_\mu\rightarrow
\overline{\nu}_e) \simeq 0.3\times 10^{-2}$, one arrives at
\be
P(\overline{\nu}_\mu\rightarrow
\overline{\nu}_e)=
0.3\times 10^{-2} = 0.02 \times \sin^{2}\left(\frac{ 1.27\times
\Delta m^2_{31}\times 30}{45}\right). \label{lsnda}
\ee
This yields,  $\Delta m^2_{3 1} = 0.2\,\,\mbox{eV}^2$. However, a
more detailed calculation that uses the LSND-$\overline\nu_\mu$
spectral function $f_{\overline\nu_\mu}$ and a cut-off at
$20\,\,\mbox{MeV}$ yields a (see Ref. [\refcite{notes}] for the
notational details)
${\cal{O}}^\prime\left(30\,\,\mbox{meters},\,\Delta m_{3 1}^2,\,
E_{min}= 20.0 \,\,\mbox{MeV}\right)= 0.15$. Fig. 2 of Ref.
[\refcite{notes}] graphs
${\cal{O}}^\prime\left(30\,\,\mbox{meters},\,\Delta m_{3 1}^2,\,
E_{min}= 20.0\,\,\mbox{MeV}\right)$
 as a function of
$\Delta m_{3 1}^2$. For
${\cal{O}}^\prime\left(30\,\,\mbox{meters},\,\Delta m_{3 1}^2,\,
E_{min}= 20.0\,\,\mbox{MeV}\right)$ below about $0.4$, this
function is single valued, and yields a unique solution. This
solution is
\be\mbox{\boldmath$
\Delta m_{3 1}^2 = 0.3 \,\,eV^2$}.\label{msd}
\ee
Taking this value for $\Delta m_{3 1}^2$ we find
$P(\overline{\nu}_e\rightarrow\overline{\nu}_e)$ for the reactor
disappearance experiments and
$P(\overline{\nu}_\mu\rightarrow\overline{\nu}_\mu)$ for $\nu_\mu$
disappearance experiments\cite{muon} read (assuming $L
\ll\lambda^{osc}_{2\,1}$):
\begin{eqnarray}
P(\overline{\nu}_e\rightarrow\overline{\nu}_e) & = & 1- 0.04
\sin^2\left(\frac{1.27\times 0.3 \times L}{E}\right), \label{reactora}\\
P(\overline{\nu}_\mu\rightarrow\overline{\nu}_\mu) & = & 1-0.998
\sin^2\left(\frac{1.27\times 0.3 \times L}{E}\right).
\end{eqnarray}
Thus,  while a detailed global analysis of existing neutrino
oscillation data is beyond the scope of this essay, the
calculations from above indicate that matrices of the type just enumerated,
(up to minor higher order corrections)
can accommodate all existing data from the
reactor-disappearance experiments,\cite{reactor} the $\nu_\mu$
disappearance experiment,\cite{muon} the LSND and KARMEN
experiments,\cite{LSND,KARMEN} and the atmospheric neutrino data
from SuperK.\cite{SK}.

\noindent
\underline{A cautionary remark:} A comparison of Eqs (\ref{lsnda})
and (\ref{reactora}) shows, for example, that a naive use of the
two dimensional $\left[\Delta m^2,
\,\sin^2(2\Theta)\right]$ exclusion plots can be quite misleading. Both  the
 two-parameter description of the $\left[\Delta m^2,
\,\sin^2(2\Theta)\right]$ formalism, and  the very specific mixing
matrix under consideration here for the five-parameter formalism,
we obtain very similar expressions for  the  neutrino-oscillation
probabilities. The amplitude of oscillation in the
$\overline{\nu}_\mu\rightarrow
\overline{\nu}_e$ channel and the
 $\overline{\nu}_e\rightarrow
\overline{\nu}_e$ channel are same for the two-parameter formalism,
while for five-parameter case, and for the specific choice of the
mixing matrix, they are different by a factor of $2$. The general
situation is often even more different.

The subject of the solar neutrino deficit  is taken up next.

\section{
Solar neutrino deficit
and the problem with a sterile neutrino solution}

Within the framework presented here,  the mixing matrix in
(\ref{ua}) yields an energy-independent solar neutrino deficit of
$0.98$. Compared to the data this number is too large by a
factor of two to three. Furthermore, the experiments rule out an
energy-independent solar neutrino deficit.\cite{jbe} The
Data/SSM-prediction ($\equiv
\delta$)\footnote{SSM
= Standard Solar Model, see Table 1 of Ref. \refcite{jbe}.}\,\, is now
known to be as follows: $\delta$
= $0.33
\pm 0.029$ (Homestake), 
$0.474\pm 0.020$ (SuperK), 
$0.52 \pm 0.06$
(SAGE),
$0.60\pm 0.06$  (GALLEX).
The Homestake chlorine experiment is the oldest running
experiment with an energy threshold of about $0.8$ MeV.
GALLEX and SAGE carry an energy threshold of about $0.2$ MeV.
SuperK,
which has a higher energy threshold of $6.5$ MeV, has the advantage
in that it is able to study the solar neutrino deficit as a
function of energy. {\bf The SuperK preliminary data carries a higher
$\mbox{\boldmath$\delta$}$  for higher neutrino energies, with $\mbox{\boldmath$\delta$}$ 
becoming
constant towards lower energies.}

With three underlying mass eigenstates one finds only two
independent mass squared differences in the standard neutrino
oscillation phenomenology. Therefore,
one cannot accommodate the energy dependence of the three length
scales implied by the existing data. Under these circumstances
one invokes a sterile neutrino to accommodate all existing data 
on neutrino oscillations.

For the vacuum-oscillation solution,
the SuperK solar neutrino data suggests an oscillation
length of the order of an astronomical unit for neutrino energies
around $10$ MeV. Thus, at lower energies of about $1$ MeV the
oscillation length becomes about $1/10$ of an astronomical unit and
the deficit becomes energy independent (because of the ``energy
averaging'' in this energy region), and the SuperK data
yields\cite{ys} $\delta
\simeq 0.375\pm 0.025$ to be compared with the $\delta$'s for the
 Homestake,  GALLEX and SAGE experiments.
Thus the solar neutrino data from SuperK comes in conflict with the 
other experiments.

One may then consider an oscillation length that fits the 
Homestake,  GALLEX, and SAGE experiments. But with such an oscillation length
it would be difficult to fit the energy dependence of the SuperK's data
on solar neutrinos.

{\bf Thus a single sterile neutrino is an unlikely candidate to accommodate
the data from the four solar neutrino experiments.}

Therefore, the scenario within the standard neutrino oscillation 
framework as
discussed above leaves the energy dependence of the solar neutrino
anomaly to be accommodated in some different manner.
One  possibility is that one of the underlying
mass eigenstates is non-relativistic. This possiblity was recently
discussed by the present author with Goldman in Ref. [\refcite{ag}].
Two other possibilities are: (a) Violation of CPT symmetry in the neutrino
sector, and/or (b) Violation of the principle of equivalence. These
are now discussed here in Secs. 6 and 7. 

\section{
Solar neutrino deficit and the possibility for  CPT symmetry violation}

In the standard model the neutrinos  and antineutrinos are
CP-conjugated partners. Therefore, the CPT symmetry requires
$\nu_\ell$ and $\overline{\nu}_\ell$ to have identical  masses.
More precisely, the CPT symmetry requires identical  expectation
values for the 
$\overline\nu_e$ and $\nu_e$
masses: $\langle m(\overline\nu_e)\rangle =
\langle m(\nu_e)\rangle $.
Moreover, these neutrinos are built from the eigenspinors  of the
charge operator in the $(1/2,0)\oplus(0,1/2)$ representation space.
Both the Dirac fields  and the Majorana fields  are expanded in
terms of the standard Dirac spinors
$\{u_\sigma(p^\mu),\,v_\sigma(p^\mu)\}$. As a result, and for
relativistic neutrinos, in both cases the helicity-flipping
oscillations appear strongly suppressed. This result is mainly due
to the orthogonality of the four spinors
$\{u_\sigma(p^\mu),\,v_\sigma(p^\mu)\}$.

In order to allow for significant  helicity-flipping oscillations
one may consider self/anti-self charge conjugate eigenspinors in
the $(1/2,0)+(0,1/2)$ representation space. In this construct  the
spinors are not orthogonal in the helicity index, but
bi-orthogonal.\cite{dvamaj}

Then referring to Eqs.(36a) and (36b) of Ref. [\refcite{dvamaj}],
and exploiting the identities given by Eqs. (48a) and (48b) there,
we immediately obtain
\be
CP\,\lambda^S(p^\mu) = \rho^A(p^{\prime \mu}),
\ee
where $p^{\prime \mu}$ is the parity transformed $p^{\mu}$, while $S$,
and $A$ refer to self- and anti-self C conjugacy of the
$(1/2,0)+(0,1/2)$ representation C-eigenspinors
$\{\lambda^S(p^\mu),\,
\rho^A(p^\mu)\}$.

The ``neutrino'' and ``antineutrino'' spinors can, therefore, be
identified  with the set $\{\lambda^S(p^\mu),\,
\rho^A(p^\mu)\}$. This raises the important possibility that
because of the {\em bi-orthogonal} nature of these spinors in the
helicity index (see Table 1 of Ref. [\refcite{dvamaj}]) helicity
flipping oscillations are the ones that are not suppressed.
However, this framework is still in its infancy and an interacting
theory has yet to be developed. One of the nice features of this
construct is that parity violation is deeply embedded in the
construct as shown by Dvoeglazov.\cite{vd}
 Nevertheless, it is to be noted that
helicity-flipping transitions can also occur in the context of
violation of the principle of equivalence and astrophysical
magnetic fields.\cite{astro} In what follows we shall assume that
helicity flipping transitions are allowed at significant level
without regard to their physical origin.

There exist various reasons to believe that gravitation plays
significant role in the `smallness' of the neutrino masses via
physics of some unification scale. If that is the case, then the
assumption of locality must be abandoned with the consequence that
CPT symmetry is no longer on a  firm theoretical
foundation.\cite{grf94,gac,ak} Within the framework outlined above,
a CPT-violating splitting of $\Delta m^2\sim 10^{-10}$ to 
$10^{-11}\,\,\mbox{eV}^2$ between the mass eigenstates underlying
$\nu_e$ and $\overline\nu_e$ can provide $\nu_e
\leftrightarrow\overline
\nu_e$ oscillations.

In this scenario, for example,  one may choose
a $\Delta m^2$ to fit the Homestake-GALLEX-SAGE data.
The reconciliation with the
the SuperK occurs because 
SuperK detects not only the  solar $\nu_e$ but
also the oscillated solar $\overline\nu_e$, whereas the three other 
experiments are sensitive only to the oscillation-reduced flux of
solar $\nu_e$. SuperK still sees the solar neutrino `deficit' in
the solar $\nu_e$-$\overline\nu_e$ flux because the detection cross
section for the  $\overline\nu_e$ is lower than that for the
$\nu_e$.

We finally present some speculative considerations on the
possibility that the principle of equivalence may be violated. We
are encouraged in these speculations by a recently discovered
incompleteness of the general relativistic description of
gravitation. The incompleteness argument is presented in  a recent
paper [\refcite{grf98}]. Further, Halprin, Leung, and Pantaleone
show that the atmospheric neutrino data as well as the  data on the
solar neutrinos point toward the same level of violation of the
principle of equivalence.

\section{Solar neutrino deficit and violation of the principle of
equivalence}

Consider a  flavor eigenstate state $\vert\nu_\ell\rangle $ given
by Eq. (\ref{1}). Let us further assume for the sake of simplicity that
each of the mass eigenstate is also an energy eigenstate,
\be
\vert\nu_\ell\rangle = \sum_{j} \,U_{\ell j}\,
\vert E_j \rangle,\quad\ell=e,\mu,\tau,
\ee
and perform a measurement of its energy. This measurement projects
the flavor eigenstate, with probability $U^\ast_{\ell j} U_{\ell
j}$ (no sum on $\ell$, or $j$), into an energy eigenstate:
\be
{\cal M_H }\,\vert\nu_\ell\rangle \mapsto \vert E_j \rangle,\quad
j=1,2,3.
\ee
Next, consider the same flavor eigenstate as before. This time a
flavor measurement is performed (without allowing the flavor state
to evolve into other flavors). Let the corresponding operator be
${\cal M_F}$. The result of measurement is (with unit
probability)\footnote{ Alternately, we could allow an evolution of
the flavor eigenstate. In that case flavor measurement projects
into one of the three flavor states with the probability given by
Eq. (\ref{eq:d})
\[
{\cal M_F}
\,\vert\nu_\ell\rangle \mapsto
\vert\nu_{\ell^\prime}\rangle,\quad\ell^\prime=e,\mu,\tau.
\]
}

\be
{\cal M_F}
\,\vert\nu_\ell\rangle \mapsto
\vert\nu_{\ell}\rangle,\quad\ell=e,\mu,\tau.
\ee

As a consequence, the measurements  $\cal M_H$ and $\cal M_F$ do
not commute while acting upon the space spanned by the flavor
eigenstates of neutrinos, $\{\nu_e,\nu_\mu,\nu_\tau,\overline
\nu_e,\overline\nu_\mu,\overline\nu_\tau\}$,
\be
\left[\cal M_H,\,\cal M_F\right] \ne 0. \label{ne}
\ee
On the other hand, for the flavor states that are mass eigenstates,
$\{e^\pm,\mu^\pm,\tau^\pm\}$, one obtains
\be
\left[\cal M_H,\,\cal M_F\right] = 0, \label{e}
\ee
because then the flavor eigenstates and energy eigenstates  are
identical.

Now the energy eigenstates considered above are simultaneously
created by the action of gravitation also on the foundation that
gravity is known to couple to matter via the energy momentum
tensor. On the other hand, the flavor states are the eigenstates
created in the electroweak interactions. In this sense, equation
(\ref{ne}) tells us that, in some quantum mechanical contexts,
gravitation and electroweak interactions have to interfere at some
more fundamental level. A gravitational measurement that projects a
system into an energy eigenstate destroys the coherence of
different mass eigenstates of a electroweak-measurement produced
flavor eigenstate. Non-commutativity expressed by equation
(\ref{ne}) indicates that flavor and gravitation are somehow
coupled to each other, and are associated with  mutually
incompatible observables.

Put into other words, in the absence of any theory that
incorporates gravitation and electroweak  interactions on an equal
footing it is reasonable to assume that gravitational interactions
of  test particles are intimately connected with the Energy
measurement operator. For states that satisfy Eq. (\ref{e})
gravitation is decoupled from flavor and one may, therefore, assert
universal gravitational coupling for all flavors. Whereas, states
for which one obtains Eq. (\ref{ne}), as is the case for neutrinos,
gravitation is not decoupled from flavor, and, one may, therefore,
not assume universal gravitational coupling for all flavors. This
then suggests that weak-interaction flavor eigenstates of neutrinos
may violate principle of equivalence, and different neutrino
flavors may couple differently to gravity. Such considerations may
underlie a conjecture of Gasperini where he postulated nonuniversal
coupling of neutrino flavors to gravity, and recently has been
considered as a serious candidate to explain the solar neutrino
anomaly.\cite{vpe,vpe2,vpe3} In fact, as already noted, Halprin,
Leung, and Pantaleone show that the atmospheric neutrino data as
well as the data on the solar neutrinos point toward the same level
of violation of the principle of equivalence.

Finally, we note the two scenarios discussed here are likely to be
deeply intertwined. The  proposed  mass squared difference 
$\Delta m^2\sim 10^{-10}$ to 
$10^{-11}\,\,\mbox{eV}^2$
for  the $\nu_e$-$\overline\nu_e$
system is perhaps only confined to electron neutrinos (and further
conjectured to be of the type introduced in Ref.
[\refcite{dvamaj}]). The muonic and tauonic neutrinos need not
belong to the construct involving $\{\lambda^S(p^\mu),
\rho^A(p^\mu)\}$. This difference may underly the specific form
of the mixing matrix $U$. The  $U$ as given in Eq. (\ref{ua})  has
a dominant block diagonal form consisting of a  $1\times 1$ matrix,
and another $2\times 2$ matrix, embedded in a $3\times 3$ matrix.
This dominant block diagonal form thus separates not only the
$\{\nu_e,\,\overline\nu_e\}$ and $\{\nu_\mu,\,\overline\nu_\mu,\,
\nu_\tau,\,\overline\nu_\tau\}$ but in the process approximately factorizes
the $\{\lambda^S(p^\mu),\,\rho^A(p^\mu)\}$ and $\{u(p^\mu),\, v
(p^\mu)\}$ degrees of freedom in the $(1/2,0)\oplus(0,1/2)$
representation space. In fact all spinors refer to a specific mass,
and for that reason it may be more appropriate to associate the
spinorial properties just indicated with the underlying mass
eigenstates rather than the flavor eigenstates.

\section{Concluding remarks}

The $L/E$-flatness of the $e$-like events, observed in the recent
atmospheric-neutrino data from SuperK, yields a severe constraint
on the neutrino mixing matrix $U$. This constraint  is expressed by
Eqs. (\ref{coupleda})-(\ref{coupledc}). When combined with other
existing experiments, Eqs. (\ref{coupleda})-(\ref{coupledc}) imply
the results given by Eqs. (\ref{ua}) and (\ref{msd}) for the
neutrino mixing matrix and the LSND relevant mass squared
difference respectively. The obtained mixing matrix is such that it
yields identical expectation values for the masses of $\nu_\mu$ and
$\nu_\tau$ neutrinos by inducing a yet-to-be-understood symmetry in
neutrino oscillations. These results are collected together for
ready reference:
\be
\mbox{\boldmath$ U \simeq
\left(\begin{array}{rrr}
 0.99 &  0.00 &   0.10 \\
 -\,0.07 &   0.71 &  0.70 \\
  -\,0.07 &    - \,0.71 &   0.70
\end{array}\right)$},\quad 
\mbox{\boldmath$\Delta  m_{LSND}^2  =  0.3
\,\, eV^2$},\label{msdp}
\ee
and 
\be
\mbox{\boldmath $\langle$}
\mbox{\boldmath $m$}
\mbox{\boldmath $($}
\mbox{\boldmath $\nu_\mu$}
\mbox{\boldmath $)$}
\mbox{\boldmath $\rangle$}
\mbox{\boldmath $=$}
\mbox{\boldmath $\langle$}
\mbox{\boldmath $m$}
\mbox{\boldmath $($}
\mbox{\boldmath $\nu_\tau$}
\mbox{\boldmath $)$}
\mbox{\boldmath $\rangle$}.\label{expp}
\ee

We considered the emerging experimental and theoretical situation
on the solar neutrino deficit and suggested that as experimental
results become more secure we may be forced into seriously
contemplating violations of CPT and/or the principle of
equivalence.

Whether the presented ``solar antineutrinos as a solution to the
solar neutrino problem'' is exercised  by Nature shall be known  in
the near future by observations of the Sudbury Neutrino Observatory
(SNO).\cite{SNO} Similarly, the results from LSND, KARMEN, SuperK,
and other neutrino oscillation experiments shall also settle the
issue as to whether or not neutrino oscillations require three
independent  mass squared differences.

Now, to close, we return again to the question asked in the opening
section of this paper, why, after the $K^0\bar K^0$ system has
Nature chosen to create another physical system whose elements are
linear superposition of different mass eigenstates? We asked, if
this may be an expression of something fundamentally new? We here
argued that this fundamentally `new something' may be the violation
of CPT symmetry, and the related violation of the principle of
equivalence. Specifically, in this scenario the mass eigenstates
underlying $\nu_e$ and $\overline\nu_e$ may carry slightly
different masses.

However, how this happens we do not know. Specifically, it may
happen, that the $\nu_\ell$ and $\overline\nu_\ell$ mixing matrices
may be slightly different. This slight difference then results in
the different expectation values for masses of $\nu_\ell$ and
$\overline\nu_\ell$ {\em without} requiring the masses of the
underlying mass eigenstates to be different. This, however, does
not lead to the required mass squared difference for the underlying
mass eigenstates. Or, it could be that underlying mass eigenstates
for $\nu_\ell$ and $\overline\nu_\ell$ carry slightly different
masses. We assumed the latter to be the case.

Another exciting possibility is that the mixing matrices for the
$\nu_\ell$ and $\overline\nu_\ell$ are slightly different, and at
the same time the underlying mass eigenstates for the $\nu_\ell$
and $\overline\nu_\ell$ also differ slightly, {\em but} with the
constraint that expectation values for the masses of $\nu_\ell$ and
$\overline\nu_\ell$ are {\em identical}. If this were to be the
case, the CPT symmetry will effectively remain unbroken -- at least,
partly, in so far as CPT symmetry requires identical
particle-antiparticle masses -- at the level of weak interactions
and at the same time provide the needed mass squared difference.
Such a scenario would provide an intriguing parallel to the result
expressed by Eq. (\ref{expp}),
\be
{\langle} m(\overline\nu_e){\rangle}= {\langle} m(\nu_e) {\rangle}.
\ee
If in the CPT-violating framework envisaged here, $\overline U$ is
to represent the mixing matrix for the $\overline\nu_\ell$, and $U$
represents the same for $\nu_\ell$, with $\overline m_j$ and $m_j$
standing for the masses of the underlying mass eigenstates,
following relations can be assumed to parallel Eq. (\ref{expp}):
\be
\langle m(\overline
\nu_\ell)\rangle\equiv
\sum_j\overline U^{\,2}_{\ell j}\, \overline m_j = \sum_j U^{\,2}
_{\ell j}\, m_j
\equiv\langle m(\nu_\ell)\rangle,\quad
\ell=e,\mu,\tau \, .
\ee

Thus, while some of our remarks remain  still speculative, the
experiments on neutrino oscillations open without doubt a window
into a new physics. The emerging experimental situation hints at a
possible violation of the  CPT symmetry and the principle of
equivalence. In support to the theory, it should be noted, that an
{\em ab initio} recent investigation of `flavor oscillation
clocks'\cite{prd_ap98} has revealed an inherent incompleteness of
the general-relativistic description of
gravitation,\cite{grf98} an
incompleteness that may carry significant implications for the
neutrino-oscillation governed physics in astrophysical
environments.\cite{vpe3,prd_ap98} In addition, as Paul Langacker
remarked at Wein'98, the SuperK results have also tested quantum
coherence for length scales upto $1.3\times 10^4\,\,\mbox{km}$ in
the atmospheric neutrino data, and to the length scales of one
astronomical unit in the solar neutrino experiment.

\vbox{
\begin{quote}
{\bibbf  ``But, irrespective of all these theoretical
considerations, one has to follow the advice of Galileo and measure
everything that can be measured.'' L. B. Okun.\cite{lo}}
\end{quote}
}

\nonumsection{Acknowledgements}

The question that I presented in the introduction is not my
question, but the one asked by  Mariana Kirchbach.  Her perpetual
stream of questions remains stimulating. My thanks, therefore.

It has also been my pleasure to interact on a daily basis with my
LSND colleagues. I extend my thanks to them.  On the theoretical
side at Los Alamos, I thank Mikkel Johnson for reading and
commenting on the manuscript. I thank Valeri Dvoeglazov for his
statesmanship in arranging a professorship at Zacatecas, and for
our ongoing conversations on the subject.

\nonumsection{References}

\end{document}